\documentclass[11pt]{article}
\usepackage{color}
\begin{document}

\def\slash#1{{\rlap{$#1$} \thinspace/}}

\begin{flushright} 
September, 2003
\end{flushright} 

\vspace{0.1cm}

\begin{Large}
\vspace{1cm}
\begin{center}
{Myers Effect and Tachyon Condensation }    \\
\end{center}
\end{Large}
\vspace{1cm}

\begin{center}
{\large  Yusuke Kimura }  \\ 

\vspace{0.5cm} 
{\it Theoretical Physics Laboratory, RIKEN, }\\
{\it Wako, Saitama 351-0198, Japan} \\

\vspace{0.5cm} 
{kimuray@postman.riken.go.jp}
\vspace{0.8cm} 

\end{center}

\vspace{1cm}

\begin{abstract}
\noindent
\end{abstract}
\noindent 
\hspace{0.4cm}

D$0$-branes are unstable in the presence of 
an R-R field strength background. 
A fuzzy two-sphere is classically stable 
under such a background, this phenomenon being 
called the Myers effect. 
We analyze this effect from the viewpoint of 
tachyon condensation. 
It is explicitly shown that a fuzzy two-sphere 
is realized by the condensation of tachyons 
which appear from strings 
connecting different D$0$-branes. 
The formation of a fuzzy $CP^{2}$ is also investigated 
by considering an $SU(3)$ invariant R-R field strength 
background. We find that 
the dynamics of the D-branes depends on the properties of 
the associated algebra. 

\newpage 



\section{Introduction}
\hspace{0.4cm}

The discovery of D-branes has led to 
exciting developments in string theory 
such as the relationship between Yang-Mills theory 
and string theory \cite{polreview, taylorreview}. 
The appreciation of noncommutative geometry 
in string theory also 
comes from developments of D-brane. 
A low energy effective action of $N$ coincident D-branes 
is described by the $U(N)$ Yang Mills theory, 
$U(N)$ adjoint scalars representing 
the transverse coordinates of this system \cite{witten}. 
Since they are given by $U(N)$ matrices, this fact suggests that 
the space-time probed by D-branes is 
related to noncommutative geometry. 
The idea of noncommutative geometry has found a 
prominent role in string theory \cite{SW}. 
Concrete models for investigating the nonperturbative aspects 
of string theory were proposed in \cite{BFSS,IKKT}, 
and the relationship between matrix models and noncommutative geometry 
were argued in \cite{CDS,AIIKKT,AMNS} for example. 

Myers showed that 
the low energy effective action of $N$ D-branes 
in the presence of a constant R-R field strength background 
has an additional Chern-Simons term, and that 
the minimum of the potential is given by 
the configuration where transverse coordinates 
form a fuzzy sphere \cite{myers}. 
The fuzzy two-sphere has been investigated from 
several points of view 
\cite{hoppethesis,madore,kabattaylor,watamura,ARS,IKTW}. 
Investigating the dynamics of curved D-branes 
is important. 
The dynamic properties of a fuzzy sphere look 
quite different from those of a flat D-brane. 
This is investigated by the matrix model action 
with the Chern-Simons term \cite{myers}. 
We can summarize that 
D$0$-branes or some fuzzy spheres 
are unstable due to the presence of an
R-R field strength background, 
and that 
only stable 
configuration is a fuzzy sphere which is given by 
the irreducible representation of $SU(2)$. 
A configuration composed of some fuzzy spheres and 
D$0$-branes 
is given by the reducible representation of $SU(2)$. 
This system has tachyonic instability, 
being considered to roll down to 
a fuzzy sphere which is given by 
the irreducible representation 
after tachyon condensation 
\cite{hashimotokrasnov,JMWY}. 
Other works investigating 
the dynamics 
of fuzzy sphere have also been reported in 
\cite{hikidanozakisuga,gaoyang,hyakutake,takatabal,
hashimoto,stabilityquantum,0305226}. 

Stories of tachyon condensation 
have been developed on various grounds 
since the conjectures made by 
Sen \cite{sen9904207}. 
Earlier works on tachyon condensation can be found in 
\cite{halpern}. 
The existence of a tachyon implies instability of 
the perturbative vacua, condensation of the tachyon 
producing a stable configuration. 
The instability indicated by 
tachyons appearing around the reducible representation 
is also considered in this way. 
The irreducible representation is expected to be realized 
after the tachyon acquires a non-vanishing 
vacuum expectation value. 
We calculate in this paper the 
tachyon potentials and explicitly show 
the process of the tachyon condensation by using 
the Yang-Mills action with the Chern-Simons term. 
We have an extra parameter which is given by 
the R-R field strength in comparison to 
the effective action in a flat background. 
By tuning 
this parameter in such a way that 
the low energy Yang-Mills description 
of this system is valid, 
we can analyze the tachyon without using 
the string field theory. 
The Yang-Mills theory provides a simple field theory model 
that enables us to analyze the process of tachyon condensation. 
There have been some studies which investigated tachyon condensation 
by using the Yang-Mills action 
\cite{ahashimototaylor,gavanarainsarmadi,
awatahiranohyaku,massartroost,aganagopaku,krausrajarashenk,mandalwadia,
hashimotonagaoka}. 

The organization of this paper is as follow. 
Section \ref{sec:myerseffectandfuzzysphere} 
presents a matrix model with the Chern-Simons 
term and classical solutions of the model. 
Fuzzy spheres and D$0$-branes appear as classical solutions 
of this model, the minimum of classical energy 
being realized by a fuzzy sphere. 
In section \ref{sec:fuzzyspherefromD0}, we explicitly show that 
the Myers effect can be viewed as the phenomenon of 
tachyon condensation. 
Tachyon potentials are obtained for some situations. 
In section \ref{sec:fuzzycp2}, we replace the 
structure constant of $SU(2)$ with that of 
$SU(3)$ in the previous matrix model. 
Not only the fuzzy sphere, but also fuzzy $CP^{2}$ 
becomes a classical solution. 
We explain the concept of fuzzy $CP^{2}$, the coordinates 
of this space being given by the generators of the 
$SU(3)$ algebra. The $(m,0)$ representation provides 
a four-dimensional space and the $(m,n)$ representation 
a six-dimensional space. 
D$0$-branes under the $SU(3)$ background are considered in 
section \ref{sec:D0branesinsu(3)}. 
An analysis analogous to that in section \ref{sec:fuzzyspherefromD0} 
is presented. 
We can observe such phenomena as the formation of 
the fuzzy two-sphere and the fuzzy $CP^{2}$. 
Moreover, interesting phenomena 
that the formation of the fuzzy $CP^{2}$ from 
D$0$-branes and fuzzy two-spheres can be seen 
due to the rich structure of the $SU(3)$ algebra. 
The stability of these curved branes constructed 
from the Lie algebra is closely related to 
the property of the associated algebra. 
Section 
\ref{sec:summaryanddiscussions} is devoted to summary and 
discussions. In Appendix \ref{sec:someformulaeofSU(3)}, 
the $(m,n)$ representation of $SU(3)$ is constructed. 
Instability of the fuzzy sphere and fuzzy $CP^{2}$ 
is shown in Appendix \ref{sec:instability} 
by considering the spectrum of open strings 
connecting a D$0$-brane and the curved branes. 


\vspace{0.6cm}

\section{Myers effect and fuzzy sphere}
\hspace{0.4cm}
\label{sec:myerseffectandfuzzysphere}
Let us first consider the effective action of 
$N$ D$0$-branes in a flat background. Such an action 
is given by the dimensional reduction to one dimension 
of $U(N)$ ten-dimensional super Yang-Mills action \cite{witten}. 
Since we are only interested in static classical solutions 
in this study, the kinetic term is ignored. 
The potential term of this action is given by 
\begin{equation}
V=\lambda^{2}T_{0}
Tr\left(-
\frac{1}{4}[\Phi_{\mu},\Phi_{\nu}][\Phi_{\mu},\Phi_{\nu}]
\right), 
\end{equation}
where $T_{0}=1/g_{s}l_{s}$ is the tension of a D$0$-brane 
and $\lambda=2\pi l_{s}^{2}$. 
The indices take values $\mu,\nu=1,\cdots,9$. 
The fields $\Phi_{\mu}$ are scalar fields describing 
transverse fluctuations of the $N$ D$0$-branes. 
Our conventions are such that $\Phi_{\mu}$ have the dimensions 
of $length^{-1}$. 
We now define fields which have the mass dimensions of $length$, 
$X_{\mu}=\lambda\Phi_{\mu}$. 
This action has two types of classical solution. 
One is $N$ D$0$-branes: 
\begin{equation}
[X_{\mu},X_{\nu}]=0,  
\end{equation}
and the other is a flat D-brane:
\begin{equation}
[X_{\mu},X_{\nu}]=-iC_{\mu\nu}{\bf 1}. 
\end{equation}
Both classical solutions are stable. 

When we consider a constant R-R four form field strength
background, 
an interesting phenomenon occurs \cite{myers}. 
The potential of the $N$ D$0$-branes has an additional 
Chern-Simons 
term\nobreak
\footnote{
We follow the notation in \cite{myers} in which 
the R-R charge is defined as $\mu_{0}=T_{0}$. 
}: 
\begin{equation}
V=\frac{T_{0}}{\lambda^{2}}Tr\left(
-\frac{1}{4}[X_{\mu},X_{\nu}][X_{\mu},X_{\nu}]
+\frac{i}{3}\alpha \epsilon_{\mu\nu\lambda}X_{\mu}[X_{\nu},X_{\lambda}]
\right). 
\label{potentialunderSU(2)}
\end{equation}
The extrema of this potential are given by 
the following equations: 
\begin{equation}
[X_{\nu}, 
\left([X_{\mu},X_{\nu}]-i\alpha \epsilon_{\mu\nu\lambda}X_{\lambda}
\right)]=0. 
\label{eofsu(2)}
\end{equation}
Classical solutions which satisfy the above equations 
are given by $N$ D$0$-branes 
\begin{equation}
[X_{\mu},X_{\nu}]=0,  
\label{D0branes}
\end{equation}
or the following non-abelian solution 
\begin{equation}
X_{\mu}=\alpha L_{\mu}, \hspace{0.4cm}
[L_{\mu},L_{\nu}]=i\epsilon_{\mu\nu\lambda}L_{\lambda}, 
\end{equation}
where $L_{\mu}$ are an $N\times N$ matrix representation of 
the $SU(2)$ algebra. 
When this representation is irreducible, 
the solution gives a spherical D$2$-brane
whose radius is given by the quadratic Casimir:
\begin{equation}
\rho^{2}=X_{\mu}X_{\mu}=
\alpha^{2}L_{\mu}^{(j)}L_{\mu}^{(j)}=\alpha^{2}j(j+1)
=\alpha^{2}(N^{2}-1)/4,  
\end{equation}
where
$L_{\mu}^{(j)}$ denote the spin 
$j=(N-1)/2$ representation of $SU(2)$. 
This solution has been investigated from the viewpoint of 
noncommutative geometry, and 
is called a fuzzy sphere or noncommutative 
sphere \cite{madore}. This fuzzy sphere is composed of 
$N$ D$0$-branes and a D$2$-brane \cite{myers}. 
A smooth sphere is obtained as $N$ 
becomes large with $\rho$ fixed. 

Reducible representations are also extrema of 
the potential. 
While the irreducible representation of 
$SU(2)$ represents a fuzzy sphere, 
the reducible representation represents some fuzzy spheres. 
Reducible representation of the form 
$L_{\mu}=diag(L_{\mu}^{(j_{1})},\cdots,L_{\mu}^{(j_{s})})$ 
denotes $s$ fuzzy spheres. 
$N$ and the radii of the fuzzy spheres are given by 
\begin{equation}
N=\sum_{r=1}^{s}N_{r}=\sum_{r=1}^{s}(2j_{r}+1),
\hspace{0.4cm}
(\rho^{2}_{r})^{2}=\alpha^{2}j_{r}(j_{r}+1). 
\end{equation}
The classical energy of the irreducible and 
reducible representations is calculated as follows, 
\begin{eqnarray}
&&E_{irr}=-\frac{\alpha^{4}}{6}j(j+1)(2j+1), 
\label{energyirr} 
\\
&&E_{red}=-\frac{\alpha^{4}}{6}
\sum_{r=1}^{s}j_{r}(j_{r}+1)(2j_{r}+1). 
\end{eqnarray}
We can find that 
the classical energy of the irreducible representation 
is less than that of the reducible representation 
when we fix $N$. Therefore, the reducible representations 
are expected to 
correspond to unstable extrema of the potential, 
collapsing into a sphere which is constructed from 
the irreducible representation. 
Instability of the reducible representations 
is closely related to the appearance of tachyonic modes 
\cite{hashimotokrasnov,JMWY}. In the next section, we explicitly check 
that an irreducible representation is realized 
by the condensation of tachyons which appear from 
off-diagonal fluctuations.  

Note that some solutions which do not belong to 
the $SU(2)$ algebra satisfy the equations of motion 
(\ref{eofsu(2)}). An example of such solutions is 
a system composed of a D$0$-brane and 
a fuzzy two-sphere: 
\begin{equation}
X_{\mu}=\alpha
\left( \begin{array}{c c}
  c_{\mu}& 0 \\  
   0 &  L_{\mu}\\
   \end{array} \right). 
\end{equation}
$c_{\mu}$ is a position of the D$0$-brane 
and $L_{\mu}$ is the spin $(N-2)/2$ representation of 
$SU(2)$. 

Although we consider the effective action of D$0$-branes, 
these analyses are not restricted to D$0$-branes. 
We may consider the effective action of D$p$-branes such that 
three transverse scalars are described 
by the action in equation (\ref{potentialunderSU(2)}). 

Before finishing this section, we estimate 
a region where these calculations are reliable. 
We use the Yang-Mills action in (\ref{potentialunderSU(2)}) 
to describe tachyons arising from 
open strings around D$0$-branes and spherical 
D$2$-branes. Using this action is valid when the commutator, 
$\lambda[\Phi_{\mu},\Phi_{\nu}]$, is small enough \cite{myers}. 
This condition 
is equivalent to $\alpha\rho =\alpha^{2}N=\rho^{2}/N 
\ll \lambda$. If we introduce 
the noncommutative scale as 
$l_{nc}^{2}=4\pi\rho^{2}/N$, the condition becomes 
\begin{equation}
l_{nc}\ll l_{s}. 
\label{lncls}
\end{equation}
It will be found that 
this condition ensures a suitably small tachyon mass 
compared to the string scale. 
Furthermore, we impose $g_{s}\ll 1$ to ensure the classical analysis.

\section{Fuzzy sphere from D$0$-branes}
\label{sec:fuzzyspherefromD0}
\hspace{0.4cm}
In this section, 
we consider a system composed of some 
D$0$-branes under a constant R-R four form 
field strength background, 
explicitly showing that 
a fuzzy two-sphere is realized by the condensation of 
tachyons which arise from strings connecting 
different D$0$-branes. 

We expand the matrices as 
\begin{equation}
X_{\mu}=\hat{x}_{\mu}+ A_{\mu}, 
\label{expansion}
\end{equation}
where $\hat{x}_{\mu}$ is a classical solution.  
In this section, the solution (\ref{D0branes}) is considered. 
$A_{\mu}$ is the fluctuation around the classical solution, 
representing the fields which appear from strings connecting 
D$0$-branes. 
Their off-diagonal components are particularly 
important, since they provide tachyonic modes. 
The potential (\ref{potentialunderSU(2)}) 
is expanded as follows: 
\begin{equation}
V=V_{0}+V_{2}+V_{3}+V_{4}, 
\end{equation}
where $V_{0}$ is the classical energy, and 
\begin{eqnarray}
&&V_{2}=\frac{1}{2}Tr\left(
A_{\nu}[\hat{x}_{\mu},[\hat{x}_{\mu},
A_{\nu}]]
\right)
+2Tr\left(A_{\mu}
([\hat{x}_{\mu},\hat{x}_{\nu}]
-i\alpha\epsilon_{\mu\nu\lambda}\hat{x}_{\lambda}
)A_{\nu}\right), \cr
&&
V_{3}=-Tr\left(
[\hat{x}_{\mu},A_{\nu}][A_{\mu},A_{\nu}]\right)
 +\frac{i}{3}\alpha \epsilon_{\mu\nu\lambda} Tr
\left(A_{\mu}[A_{\nu},A_{\lambda}]\right) \cr
&&
V_{4}=-\frac{1}{4}Tr\left(
[A_{\mu},A_{\nu}][A_{\mu},A_{\nu}]\right). 
\end{eqnarray}
$V_{2}$ gives mass terms, and $V_{3}$ and $V_{4}$ give 
interaction terms. 
Note that we added the following term to fix the $U(N)$ gauge invariance, 
\begin{equation}
-\frac{1}{2}Tr([\hat{x}_{\mu},A_{\mu}][\hat{x}^{\nu},A^{\nu}]). 
\end{equation}
The overall factor in front of the potential 
(\ref{potentialunderSU(2)}) has been ignored since 
it is not important in the present discussion. 
We first show that tachyons appear from the off-diagonal 
components of the fluctuations. 
We then calculate $V_{3}$ and $V_{4}$ for the tachyonic fields and 
search for a stable extremum in which the tachyonic 
fields have non-zero expectation values. 
Finally, 
$X_{\mu}$ can be expected to satisfy the algebra of 
a fuzzy two-sphere.   
Similar calculations for brane and anti-brane system 
have been presented in \cite{krausrajarashenk,mandalwadia}. 

\vspace{0.8cm}

We first study a system of two D$0$-branes 
as the simplest case. 
We choose the positions of the D$0$-branes as follows: 
\begin{equation}
\hat{x}_{1}=\hat{x}_{2}=0, \hspace{0.4cm}
\hat{x}_{3}=
\left( \begin{array}{c c}
  x_{3}^{(1)}&0  \\  
  0 &x_{3}^{(2)}   \\
   \end{array} \right)
\hspace{0.4cm}
(x_{3}^{(1)} \ge x_{3}^{(2)}).
\end{equation}
This choice is without loss of
generality. 
Since the potential (\ref{potentialunderSU(2)}) has the translation 
invariance 
\begin{equation}
\delta X_{\mu}=c_{\mu}{\bf 1}_{N}, 
\end{equation}
we restrict $X_{\mu}$ to be traceless 
by assuming the condition 
$x_{3}^{(1)}+x_{3}^{(2)}=0$. 
Fields appearing in this system are introduced 
as 
\begin{equation}
A_{\mu}=
\left( \begin{array}{c c c}
  0&a_{\mu}  \\  
  \bar{a}_{\mu} & 0   \\
   \end{array} \right), 
\end{equation}
where 
$\bar{a}$ is a complex conjugate of $a$. 
The off-diagonal components represent 
fields which appear from strings connecting 
different D-branes, and are considered to 
play an important role in noncommutative configurations. 
We now investigate a situation where noncommutative 
configurations are constructed after tachyons 
appearing from the off-diagonal parts condense. 
We do not turn on the diagonal components 
since they do not give tachyonic modes. 
These components are actually set to zero 
by using the equations of motion for them. 
Although there may be possibility that the diagonal 
comonents can have nonzero values due to 
interactions with off-diagonal fields, 
we first assume that the diagonal parts do not 
contribute to noncommutative solutions to simplify 
the calculation. 
A noncommutative (fuzzy sphere) solution is actually obtained 
by considering only off-diagonal parts.

We can easily calculate the mass term of 
the off-diagonal components as follows, 
\begin{eqnarray}
V_{2}&=&a_{\mu}(\delta_{\mu\nu}\Delta^{2}
+2i\alpha\epsilon_{\mu\nu3}\Delta
)\bar{a}_{\nu}\cr
&=&(\Delta^{2}-2\alpha\Delta)a_{+}\bar{a}_{-}
+(\Delta^{2}+2\alpha\Delta)a_{-}\bar{a}_{+}
+\Delta^{2}a_{3}\bar{a}_{3} 
\end{eqnarray}
where $\Delta \equiv x_{3}^{(1)}-x_{3}^{(2)} \ge 0$ and 
\begin{equation}
a_{+} \equiv \frac{1}{\sqrt{2}}(a_{1}+ia_{2}),
\hspace{0.4cm}
\bar{a}_{\pm} \equiv \frac{1}{\sqrt{2}}(\bar{a}_{1}\pm i\bar{a}_{2})
\label{a+-}
\end{equation}
Note that $\bar{a}_{\mp}$ are complex conjugate of $a_{\pm}$. 
These fields are complex, since the strings connecting 
different D-branes are orientable. 
Since $\Delta$ is positive, 
only $a_{+}$ can be tachyonic. 
A complex tachyon appears when the distance between 
two D-branes is short, $0< \Delta< 2\alpha$. 
We are interested in this case. 
Non-tachyonic modes are set to zero 
$a_{-}=a_{3}=0$ (by using the equations of motion), 
and we parameterize $a_{+}$ as 
\begin{equation}
a_{+}=te^{i\theta}, 
\end{equation}
where $t \ge 0$ and $0 \le \theta < 2\pi$. 
The full potential 
\footnote{The interaction term $V_{3}$
is zero in the present situations, while 
it is not zero in section \ref{sec:D0branesinsu(3)}. 
} 
is calculated as 
\begin{equation}
V=(\Delta^{2}-2\alpha\Delta)t^{2}+t^{4}
\equiv m^{2}t^{2}+t^{4}. 
\end{equation}
This potential does not depend on $\theta$. 
This fact reflects the $U(1)$ symmetry of this system 
which is related to the rotation 
around the $3$-axis.  
We look for extrema of this potential 
which satisfy $dV/dt=0$. 
It should be noted that condition 
$dV/dt=0$ is equivalent to the condition that 
$X_{\mu}$ satisfy the equations of motion (\ref{eofsu(2)}). 
The point $t=0$ corresponding to 
two D$0$-branes is an unstable extremum. 
The value of $t$ corresponding to a stable extremum is 
found to be 
\begin{equation}
t^{2}=-\frac{1}{2}m^{2}. 
\end{equation}
At this point, $V$ takes the following value, 
\begin{equation}
V=-\frac{1}{4}m^{4}=-\frac{1}{4}
\left( \Delta^{2}-2\alpha \Delta \right)^{2},
\end{equation}
which is lower than the classical potential energy of 
two D$0$-branes. 
We next look for a configuration which minimizes this potential. 
It is minimized when $\Delta=\alpha$. 
In this case, the tachyon mass and the classical energy 
of this configuration are 
$m^{2}=-\alpha^{2}$ and 
$V=-\alpha^{4}/4$. 
This value of $V$ is indeed the classical energy of 
a fuzzy sphere which is constructed from 
the $j=1/2$ representation of $SU(2)$ 
(see (\ref{energyirr})). 
We can easily recognize that matrices 
$X_{\mu}$ can be written as 
$X_{\mu}=\alpha \sigma_{\mu}/2$
\footnote{ 
We have set $\theta =0$ in this expression. 
}. 

The tachyon mass is given by 
\begin{equation}
m_{t}^{2}=-\frac{\alpha^{2}}{l_{s}^{4}}=
-\frac{l_{nc}^{2}}{N l_{s}^{2}}\frac{1}{l_{s}^{2}}
\end{equation}
after restoring the string scale. 
The restriction (\ref{lncls}) ensures that 
this tachyon mass is much smaller than the string mass scale. 

Let us now investigate what configuration 
is realized when the tachyon 
condenses to another value. 
If we fix the distance between two D$0$-branes as 
$\Delta =\alpha/2$ for example, 
the tachyon potential becomes $V=-\frac{3}{4}\alpha^{2}t^{2}+t^{4}$. 
Then the tachyon can condense at $t^{2}=3\alpha^{2}/8$, 
realizing the following solution: 
\begin{equation}
X_{1}=\frac{\sqrt{3}}{4}\alpha\sigma_{1},
\hspace{0.2cm}
X_{2}=\frac{\sqrt{3}}{4}\alpha\sigma_{2},
\hspace{0.2cm}
X_{3}=\frac{\alpha}{4}\sigma_{3}. 
\label{ellipsoidal}
\end{equation}
A fuzzy ellipsoidal sphere is realized. 
The energy of this solution is $V=-9\alpha^{4}/64$, 
which is higher than that of a fuzzy two-sphere.  
By comparing this solution with a fuzzy two-sphere solution, 
we find that a solution which has higher symmetry has 
lower energy. 

\vspace{0.8cm}

We next argue a system of three D$0$-branes whose positions are 
\begin{equation}
\hat{x}_{1}=\hat{x}_{2}=0, \hspace{0.4cm}
\hat{x}_{3}=
\left( \begin{array}{c c c}
  x_{3}^{(1)}&0 &0 \\  
  0 &x_{3}^{(2)}   & 0 \\
  0&0 & x_{3}^{(3)} \\
   \end{array} \right). 
\end{equation}
$x_{3}^{(i)}$ are restricted to 
$x_{3}^{(1)} \ge x_{3}^{(2)} \ge x_{3}^{(3)}$ 
and $x_{3}^{(1)} + x_{3}^{(2)} + x_{3}^{(3)}=0$ 
without loss of
generality. 
We denote fluctuations by 
\begin{equation}
A_{\mu}=
\left( \begin{array}{c c c}
  0&a_{\mu} &b_{\mu} \\  
  \bar{a}_{\mu} & 0  & c_{\mu} \\
  \bar{b}_{\mu} &\ \bar{c}_{\mu} & 0 \\
   \end{array} \right). 
\end{equation}
The calculation is analogous to the previous case. 
The mass terms of these fields are 
\begin{eqnarray}
V_{2}&=&a_{\mu}(\delta_{\mu\nu}\Delta_{12}^{2}
+2i\alpha\epsilon_{\mu\nu3}\Delta_{12}
)\bar{a}_{\nu}\cr
&&+b_{\mu}(\delta_{\mu\nu}\Delta_{13}^{2}
+2i\alpha\epsilon_{\mu\nu3}\Delta_{13}
)\bar{b}_{\nu}
+c_{\mu}(\delta_{\mu\nu}\Delta_{23}^{2}
+2i\alpha\epsilon_{\mu\nu3}\Delta_{23}
)\bar{c}_{\nu} \cr
&=&(\Delta_{12}^{2}-2\alpha\Delta_{12})a_{+}\bar{a}_{-}
+(\Delta_{12}^{2}+2\alpha\Delta_{12})a_{-}\bar{a}_{+}
+\Delta_{12}^{2}a_{3}\bar{a}_{3} \cr
&&+(\Delta_{13}^{2}-2\alpha\Delta_{13})b_{+}\bar{b}_{-}
+(\Delta_{13}^{2}+2\alpha\Delta_{13})b_{-}\bar{b}_{+}
+\Delta_{13}^{2}b_{3}\bar{b}_{3} \cr
&&+(\Delta_{23}^{2}-2\alpha\Delta_{23})c_{+}\bar{c}_{-}
+(\Delta_{23}^{2}+2\alpha\Delta_{23})c_{-}\bar{c}_{+}
+\Delta_{23}^{2}c_{3}\bar{c}_{3},
\end{eqnarray}
where $\Delta_{ij}=x_{3}^{(i)}-x_{3}^{(j)}$.  
$a_{\pm}$ are defined in (\ref{a+-}), 
$b_{\pm}$ and $c_{\pm}$ being analogously defined. 
Since $\Delta$ is positive, three complex tachyon fields can appear 
when $0 < \Delta < 2\alpha$. 
After dropping the fields except $a_{+}$, $b_{+}$ and $c_{+}$, 
$V_{4}$ can be evaluated as follows, 
\begin{eqnarray}
V_{4}
&=&(a_{+}\bar{a}_{-})^{2}+(b_{+}\bar{b}_{-})^{2}
+(c_{+}\bar{c}_{-})^{2} \cr
&&-(a_{+}\bar{a}_{-})(c_{+}\bar{c}_{-})
+2(a_{+}\bar{a}_{-})(b_{+}\bar{b}_{-})
+2(b_{+}\bar{b}_{-})(c_{+}\bar{c}_{-}). 
\end{eqnarray}
Then the full potential for those fields 
takes the following form
\begin{eqnarray}
V=m_{1}^{2}t_{1}^{2}+m_{2}^{2}t_{2}^{2}+m_{3}^{2}t_{3}^{2}
+t_{1}^{4}+t_{2}^{4}+t_{3}^{4}
-t_{1}^{2}t_{2}^{2}+2t_{1}^{2}t_{3}^{2}+2t_{2}^{2}t_{3}^{2}, 
\label{N=3fullpotential}
\end{eqnarray}
where $t_{i}$ and $m_{i}$ are defined as 
\begin{equation}
a_{+}=t_{1}e^{i\theta_{1}}, \hspace{0.4cm}
c_{+}=t_{2}e^{i\theta_{2}}, \hspace{0.4cm}
b_{+}=t_{3}e^{i\theta_{3}},
\end{equation}
and 
\begin{equation}
m_{1}=\Delta_{12}^{2}-2\alpha\Delta_{12},\hspace{0.4cm} 
m_{2}=\Delta_{23}^{2}-2\alpha\Delta_{23},\hspace{0.4cm}
m_{3}=\Delta_{13}^{2}-2\alpha\Delta_{13}. 
\label{massoftN=3SU(2)}
\end{equation}
This is also independent of $\theta$, reflecting 
the $U(1)$ symmetry. 
The point corresponding to the condensation of tachyons 
is given 
by a stable extremum of the potential. 
This can be found by solving the following equations, 
\begin{eqnarray}
&&\frac{d V}{d t_{1}}
=(m_{1}^{2}+2t_{1}^{2}-t_{2}^{2}+2t_{3}^{2})t_{1}=0, \cr
&&\frac{d V}{d t_{2}}
=(m_{2}^{2}+2t_{2}^{2}-t_{1}^{2}+2t_{3}^{2})t_{2}=0, \cr
&&\frac{d V}{d t_{3}}
=(m_{3}^{2}+2t_{3}^{2}+2t_{1}^{2}+2t_{2}^{2})t_{3}=0. 
\label{eomfort}
\end{eqnarray}
The point 
$t_{1}=t_{2}=t_{3}=0$ is a trivial solution, which represents 
an unstable configuration. 
We have to search for a point corresponding to a local minimum. 
Since solutions of these equations 
depend on the values of $\Delta_{12}$, 
$\Delta_{23}$ and $\Delta_{13}$, 
the problem is complicated. 
We therefore consider the following situation 
to simplify the analysis, 
\begin{equation}
\Delta_{23}=\alpha, \hspace{0.4cm}
\Delta_{12}\equiv \Delta >0, \hspace{0.4cm}
\Delta_{13}= \Delta+\alpha. 
\end{equation}
The mass (\ref{massoftN=3SU(2)}) can then be rewritten in terms 
of $\Delta$ as 
\begin{equation}
m_{1}^{2}=\Delta(\Delta-2\alpha), 
\hspace{0.2cm} 
m_{2}^{2}=-\alpha^{2}, 
\hspace{0.2cm}
m_{3}^{2}=(\Delta+\alpha)(\Delta-\alpha). 
\label{massintermofdelta}
\end{equation}
$m_{2}^{2}$ is always negative, and $m_{1}^{2}$ and $m_{3}^{2}$ 
can be negative when $0<\Delta <2\alpha$ and 
$0<\Delta <\alpha$ respectively. 
We have to consider 
the following three cases:  
\begin{center}
\begin{tabular}{c c }
I:& $0 < \Delta < \alpha$, \\   
\end{tabular}
\begin{tabular}{c c }
II:& $\alpha \le \Delta < \Delta_{0}$, \\  
\end{tabular}
\begin{tabular}{c c }
III:& $ \Delta_{0} \le \Delta $. \\ 
\end{tabular}
\end{center}
It will be clear soon why we have introduced $\Delta_{0}$. 
Let us first consider the situation where 
$\Delta$ is large enough (case III). 
The definition of $\Delta_{0}$ is such that only $t_{2}$ 
is tachyonic in the case III. 
The tachyon potential is given by 
\begin{equation}
V=-\alpha^{2}t_{2}^{2}+t_{2}^{4}. 
\label{potentialIII}
\end{equation}
Note that 
this potential  
does not depend on $\Delta$. 
If we change the position of the first D$0$-brane 
such that $\Delta \ge  \Delta_{0}$, 
the form of this potential does not change. 
At the minimum of this potential
$t_{2}^{2}=\alpha^{2}/2$, 
the following configuration is realized: 
\begin{equation}
X_{1}=
\left( \begin{array}{c c}
  0&  \\  
   & \frac{\alpha}{2}\sigma_{1}  \\
   \end{array} \right),  \hspace{0.2cm}
X_{2}=
\left( \begin{array}{c c}
  0&  \\  
   & \frac{\alpha}{2}\sigma_{2}  \\
   \end{array} \right),\hspace{0.2cm}
X_{3}=
\left( \begin{array}{c c}
   c &  \\  
   & \frac{\alpha}{2}\sigma_{3}-d 1_{2}  \\
   \end{array} \right).
\label{D0+D2}
\end{equation}
where 
$c=\frac{\alpha}{3}+\frac{2}{3}\Delta$
and $d=\frac{\alpha}{6}+\frac{1}{3}\Delta$. 
A D$0$-brane is in the position $x_{3}=c$, while a spherical 
D$2$-brane which is given by the spin-$1/2$ representation of 
$SU(2)$ is in the position $x_{3}=-d$. 
Though this matrix $X_{\mu}$ does not satisfy the SU(2) algebra, 
it is a solution of the equations of motion (\ref{eofsu(2)}). 

We next consider case II, and determine the value of $\Delta_{0}$. 
Since $m_{2}^{2}$ is always negative, $t_{2}$ can always take 
a non-zero solution in (\ref{eomfort}). 
On the other hand, 
whether $t_{1}$ can take a non-zero value 
or not depends on $\Delta$. Naively, $t_{1}$ is zero 
in a case $\Delta \ge 2\alpha$ because 
$m_{1}^{2}$ is positive. We must draw attention to 
the existence of the interaction term $-t_{1}^{2}t_{2}^{2}$ 
in (\ref{N=3fullpotential}). 
$\Delta_{0}$ is not $2\alpha$ due to this term 
($\Delta_{0}$ is larger than $2\alpha$). 
We now require that both $t_{1}$ and $t_{2}$ can condense to 
non-zero values. 
The potential for them is 
\begin{equation}
V=m_{1}^{2}t_{1}^{2}+t_{1}^{4}
-\alpha^{2}t_{2}^{2}+t_{2}^{4}
-t_{1}^{2}t_{2}^{2}. 
\label{tachyonpotentialinregionII}
\end{equation}
From the 
equations of motion (\ref{eomfort}), 
tachyons $t_{1}$ and $t_{2}$ respectively can condense at 
the following values, 
\begin{eqnarray}
&&t_{1}^{2}=-\frac{2}{3}m_{1}^{2}-\frac{1}{3}m_{2}^{2}
=-\frac{2}{3}(\Delta - \alpha)^{2}+\alpha^{2}
\cr
&&t_{2}^{2}=-\frac{1}{3}m_{1}^{2}-\frac{2}{3}m_{2}^{2}
=-\frac{1}{3}(\Delta - \alpha)^{2}+\alpha^{2}. 
\end{eqnarray}
$\Delta_{0}$ is determined by the condition that 
both $t_{1}^{2}$ and $t_{2}^{2}$ have to be positive. 
The conditions $t_{1}^{2} >0$ and $t_{2}^{2} >0$ provide 
$0<\Delta <\left(1+\sqrt{\frac{3}{2}}\right)\alpha$ 
and $0<\Delta <(1+\sqrt{3})\alpha$ respectively. 
Therefore $\Delta_{0}$ needs to be 
$\left(1+\sqrt{\frac{3}{2}}\right)\alpha$. 
We then substitute $t_{1}^{2}$ and $t_{2}^{2}$ into $V$. 
It can be found, after some calculations, 
that $V$ is minimized when $\Delta=\alpha$, that is  
\begin{equation}
t_{1}=\alpha, 
\hspace{0.4cm} 
t_{2}=\alpha, 
\hspace{0.4cm}
t_{3}=0.
\end{equation}
The value of $V$ is 
\begin{equation}
V=-\alpha^{4}.
\end{equation}
It can be easily checked that 
$X_{\mu}$ satisfy the spin-$1$ representation 
of $SU(2)$. A fuzzy sphere is obtained. 

In case I, three fields are tachyonic, which 
can condense to the following 
values, 
\begin{equation}
t_{1}^{2}=-\frac{2}{3}m_{1}^{2}-\frac{1}{3}m_{2}^{2}, 
\hspace{0.4cm}
t_{2}^{2}=-\frac{1}{3}m_{1}^{2}-\frac{2}{3}m_{2}^{2}, 
\hspace{0.4cm}
t_{3}=0. 
\label{seemunstable}
\end{equation}
where $m_{1}^{2}$ and $m_{2}^{2}$ are given by (\ref{massintermofdelta}). 
This point seems to be unstable, since $t_{3}$ is zero. 
It is, however, stable since 
the interaction terms $\langle t_{1}^{2}\rangle t_{3}^{2}$ and 
$\langle t_{2}^{2} \rangle t_{3}^{2}$ 
in the potential (\ref{N=3fullpotential}) 
can give the mass of $t_{3}$. 
The stability of this point can be easily confirmed as 
$d^{2}V/2dt_{1}^{2}
=-\frac{8}{3}m_{1}^{2}-\frac{4}{3}m_{2}^{2}>0$, 
$d^{2}V/2dt_{2}^{2}
=-\frac{4}{3}m_{1}^{2}-\frac{8}{3}m_{2}^{2}>0$ and 
$d^{2}V/2dt_{3}^{2}
=m_{3}^{2}-2m_{1}^{2}-2m_{2}^{2}>0$. 
The configuration that is realized in this case is 
one which is similar to 
a fuzzy ellipsoidal sphere (\ref{ellipsoidal}). 

\vspace{0.5cm}

It is worth while examining the D$0$-D$2$ system (\ref{D0+D2}) 
more closely. This system is stable whenever the position of 
the D$0$-brane is $\Delta\ge \Delta_{0}$. 
As has already been 
investigated \cite{hashimotokrasnov,JMWY}, 
this system 
becomes unstable when the D$0$-brane is close to the surface of 
the spherical D$2$-brane. 
From equation (\ref{unstableregion}), 
a tachyonic mode appears 
when $\Delta$ takes the following interval,  
\begin{equation}
\frac{3}{2}-\sqrt{\frac{3}{2}} < \frac{1}{2}+\frac{\Delta}{\alpha}
<\frac{3}{2}+\sqrt{\frac{3}{2}}. 
\label{tachyonicregionofdelta}
\end{equation}
The upper bound of $\Delta$ is equivalent to $\Delta_{0}$. 
This system is stable when the D$0$-brane 
is far enough from the D$2$-brane. 
As we decrease the distance between 
the D$0$-brane and the D$2$-brane, the system becomes unstable 
and a fuzzy sphere of larger radius is formed. 
We will explicitly show this phenomenon from the viewpoint of 
tachyon condensation. 

The D$0$-D$2$ system is unstable when $\Delta$ satisfies 
(\ref{tachyonicregionofdelta}), and 
the tachyon arising from this system can be analyzed by 
considering the tachyon potential shown in 
(\ref{tachyonpotentialinregionII}). 
The point specified by $t_{1}=0$ and $t_{2}=\alpha/\sqrt{2}$ 
is an unstable extremum, 
the configuration expressed in (\ref{D0+D2}) being realized. 
Let us expand the potential around this point, 
\begin{equation}
t_{1}=0+v, \hspace{0.4cm} 
t_{2}=\frac{\alpha}{\sqrt{2}}+a. 
\end{equation}
Fields 
$v$ and $a$ represent fluctuations around 
the expression in (\ref{D0+D2}):  
\begin{equation}
X_{\mu}=
\left( \begin{array}{c c }
   c_{\mu} &  \\  
   & \frac{\alpha}{2}\sigma_{\mu}-d_{\mu} 1_{2}  \\
   \end{array} \right)
+\left( \begin{array}{c c }
   0 &  v_{\mu} \\  
   v_{\mu}^{\dagger} & a_{\mu}  \\
   \end{array} \right).
\label{fluctuationaroundD0D2}
\end{equation}
where $v_{1}=\left(\frac{1}{\sqrt{2}}v,0\right)$, 
$v_{2}=\left(\frac{-i}{\sqrt{2}}v,0\right)$, 
$a_{1}=\frac{1}{\sqrt{2}}a \sigma_{1}$ and 
$a_{2}=\frac{1}{\sqrt{2}}a \sigma_{2}$. $a_{\mu}$ is a gauge field 
of the noncommutative gauge theory on the fuzzy sphere \cite{IKTW}. 
The potential (\ref{tachyonpotentialinregionII}) 
is expressed in terms of $v$ and $a$ as follows, 
\begin{equation}
V=-\frac{\alpha^{4}}{4}+2\alpha^{2}a^{2}
+2\sqrt{2}\alpha a^{3}+a^{4}+
\left(m_{1}^{2}-\frac{\alpha^{2}}{2}\right)v^{2}+v^{4}-\sqrt{2}\alpha a v^{2}
-v^{2}a^{2}. 
\label{potentialvanda}
\end{equation}
The mass term for $v$ is $(\Delta^{2}-2\alpha\Delta-\alpha^{2}/2$) and 
becomes negative when $\Delta$ satisfies 
(\ref{tachyonicregionofdelta}). 
Although the mass of $a$ is positive, 
it can be negative as will be shown later 
because of the interaction term $-v^{2}a^{2}$. 
The condition $d V/d v=0$ determines the extrema: 
\begin{equation}
(4v^{2}-2a^{2}-2\sqrt{2}\alpha a
-3\alpha^{2})v=0. 
\end{equation}
Since we want an extremum where $v$ takes a non-zero value, 
we substitute $4v^{2}=2a^{2}+2\sqrt{2}\alpha a
+3\alpha^{2}$ into (\ref{potentialvanda}). 
Finally we get the following potential for $a$: 
\begin{equation}
V=-\frac{13}{16}\alpha^{4}-\frac{3\sqrt{2}}{4}\alpha^{3}a
+\frac{3}{4}\alpha^{2}a^{2}+\frac{3\sqrt{2}}{2}\alpha a^{3}
+\frac{3}{4}a^{4}. 
\end{equation}
The extrema of this potential are obtained from $d V/d a=0$ as 
\begin{eqnarray}
3\left(\frac{a}{\alpha}+1+\frac{1}{\sqrt{2}}\right)
\left(\frac{a}{\alpha}+\frac{1}{\sqrt{2}}\right)
\left(\frac{a}{\alpha}-1+\frac{1}{\sqrt{2}}\right)=0. 
\end{eqnarray}
The solution $a=\left(1-1/\sqrt{2}\right)\alpha$ 
is what we are looking for. 
$v$ is determined as $v=1$. 
Substituting these values into equation (\ref{fluctuationaroundD0D2}), 
$X_{\mu}$ provide the coordinates of the fuzzy sphere 
which is given by 
the spin-$1$ representation of $SU(2)$. 
We have thus shown that 
a system comprising a D$0$-brane and a fuzzy sphere which is given by 
the spin-$1/2$ representation collapses into 
a fuzzy sphere which is given by the spin-$1$ representation 
after the condensation of tachyon appearing from string 
connecting the D$0$- and D$2$-brane.

\vspace{0.8cm}

A system composed of 
$N$ D$0$-branes can be also analogously analyzed. 
To simplify the calculation, we start from the 
following configuration: 
\begin{equation}
\hat{x}_{1}=\hat{x}_{2}=0, \hspace{0.4cm}
\hat{x}_{3}=
\left( \begin{array}{c c c c}
  x_{3}^{(1)}&  &  &\\  
   &x_{3}^{(2)}   &  & \\
  & &  \ddots &  \\
  & &  &x_{3}^{(N)}\\
   \end{array} \right)
\end{equation}
where $\Delta_{ij}=x_{3}^{(i)}-x_{3}^{(j)} \ge 0$. 
It is sufficient to consider the following fluctuations,  
\begin{equation}
A_{+}=
\left( \begin{array}{c c c c}
  0& a_{+}^{(1)} &  &\\  
   &0   & a_{+}^{(2)} & \\
  & &  \ddots &  a_{+}^{(N-1)}\\
  & &  & 0 \\
   \end{array} \right), \hspace{0.4cm}
A_{-}=
\left( \begin{array}{c c c c}
  0&  &  &\\  
   \bar{a}_{-}^{(1)}&0   &  & \\
  & \bar{a}_{-}^{(2)}&  \ddots &  \\
  & &  \bar{a}_{-}^{(N-1)} & 0 \\
   \end{array} \right). 
\end{equation}
The mass terms are 
\begin{equation}
V_{2}=m_{1}^{2}(a_{+}^{(1)}\bar{a}_{-}^{(1)})
+m_{2}^{2}(a_{+}^{(2)}\bar{a}_{-}^{(2)})
+\cdots
+m_{1}^{N-1}(a_{+}^{(N-1)}\bar{a}_{-}^{(N-1)}), 
\end{equation}
where 
$m_{i}^{2}=\Delta_{i,i+1}(\Delta_{i,i+1}-2\alpha)$, 
and interaction terms are 
\begin{eqnarray}
V_{4}&=&2(a_{+}^{(1)}\bar{a}_{-}^{(1)})^{2}
+\cdots+2(a_{+}^{(N-1)}\bar{a}_{-}^{(N-1)})^{2} \cr
&&-2(a_{+}^{(1)}\bar{a}_{-}^{(1)})(a_{+}^{(2)}\bar{a}_{-}^{(2)})
-2(a_{+}^{(2)}\bar{a}_{-}^{(2)})(a_{+}^{(3)}\bar{a}_{-}^{(3)})
\cdots
-2
(a_{+}^{(N-2)}\bar{a}_{-}^{(N-2)})(a_{+}^{(N-1)}\bar{a}_{-}^{(N-1)}). 
\end{eqnarray}
We then have 
\begin{eqnarray}
V=m_{1}^{2}t_{1}^{2}+\cdots +m_{N-1}^{2}t_{N-1}^{2} 
+t_{1}^{4}+\cdots+t_{n-1}^{4}
-(t_{1}^{2}t_{2}^{2}+\cdots+t_{N-2}^{2}t_{N-1}^{2}). 
\end{eqnarray}
Minimizing the tachyon potential gives 
$\Delta_{i,i+1}=\alpha$ for all $i$, and 
\begin{equation}
t_{1}=\frac{\alpha}{2}f(j-1), \hspace{0.2cm}
t_{2}=\frac{\alpha}{2}f(j-2), \ldots 
t_{N-1}=\frac{\alpha}{2}f(-j),
\end{equation}
where $f(m)=\sqrt{j(j+1)-m(m+1)})$. 
This solution surely provides 
the spin $(N-1)/2$ representation of $SU(2)$.


\section{Fuzzy $CP^{2}$}
\label{sec:fuzzycp2}
\hspace{0.4cm}
We investigated in the previous section the process for 
the formation of a fuzzy sphere 
under the $SU(2)$ invariant R-R four form background 
from the viewpoint of tachyon condensation. 
The reducible representations of $SU(2)$ (and 
D$0$-branes) are basically unstable, forming the irreducible 
representation of SU(2) by the condensation of tachyons 
which arise from strings between the different D-branes. 
In this section, we consider what phenomena happen 
when we replace the structure constant of $SU(2)$ 
with that of $SU(3)$. Such matrix models have previously been 
investigated in \cite{trivedivaidya,kitazawa}. 
Since the structure of $SU(3)$ is richer than 
that of $SU(2)$, in addition to the fact that 
the $SU(3)$ algebra contains the $SU(2)$ algebra as 
subalgebra, the dynamics 
of D-branes under this background becomes 
more complicated than that in the previous section. 

The potential we will consider is as follows: 
\begin{equation}
V=\frac{T_{0}}{\lambda^{2}}Tr\left(
-\frac{1}{4}[X_{\mu},X_{\nu}][X_{\mu},X_{\nu}]
+\frac{i}{3}\alpha f_{\mu\nu\lambda}X_{\mu}[X_{\nu},X_{\lambda}]
\right)\hspace{0.4cm}(\mu,\nu =1,\ldots,8). 
\end{equation}
Since we need eight transverse coordinates, this can be realized 
only in D$p$$(p\le 1)$-branes. 
Static classical solutions  of this action are 
given by the following equations, 
\begin{equation}
[X_{\nu}, 
\left([X_{\mu},X_{\nu}]-i\alpha f_{\mu\nu\lambda}X_{\lambda}
\right)]=0. 
\end{equation}
The main classical solutions involve $N$ D$0$-branes
\begin{equation}
[X_{\mu},X_{\nu}]=0, 
\end{equation}
fuzzy two-sphere 
\begin{eqnarray}
&&[X_{\mu},X_{\nu}]=i\alpha\epsilon_{\mu\nu\lambda}X_{\lambda}
\hspace{0.4cm}(\mu,\nu =1,\ldots,3), 
\end{eqnarray}
and fuzzy $CP^{2}$
\begin{equation}
[X_{\mu},X_{\nu}]=i\alpha f_{\mu\nu\lambda}X_{\lambda}
\hspace{0.4cm}(\mu,\nu =1,\ldots,8). 
\label{su(3)space} 
\end{equation}
Before studying the dynamics of these branes, 
we will study the last solution. 

The coordinates of fuzzy $CP^{2}$ are provided by 
\begin{equation}
\hat{x}_{\mu}=\alpha T_{\mu}^{(m,n)}, 
\end{equation}
where $T_{\mu}^{(m,n)}$ are the generators of 
the $(m,n)$ representation of 
$SU(3)$. 
They are realized by matrices whose size is 
\begin{equation}
N=\frac{1}{2}(m+1)(n+1)(m+n+2). 
\label{sizesu(3)matrix}
\end{equation}
Therefore the value of $N$ is restricted 
to realize this noncommutative 
solution, which is not the case in the fuzzy two-sphere. 
There have been some reports 
\cite{nairrandjbar,trivedivaidya,ABIY,ydri,
karabalinair,alexanian,kitazawa,pawelczyk0203110} 
investigating such a noncommutative space. 
The radius of this space is given by using 
the formula (\ref{casimirsu3}), 
\begin{equation}
\rho^{2}=\hat{x}_{\mu}\hat{x}_{\mu}
=\alpha ^{2}T_{\mu}T_{\mu}=\alpha ^{2}C_{2}(m,n),
\end{equation}
This space has $SU(3)$ isometry. 
There are two manifolds whose isometry is 
$SU(3)$, that is, $SU(3)/U(2)$ and 
$SU(3)/U(1)\times U(1)$. 
The noncommutative space which is 
related to the $(m,0)$ or $(0,n)$ representation 
of $SU(3)$ is $SU(3)/U(2)=CP^{2}$, while 
the noncommutative space which is 
related to the $(m,m)$ representation 
of $SU(3)$ is $SU(3)/U(1)\times U(1)$. 
The latter space is locally $CP^{2}\times S^{2}$. 
$T_{8}$ is usually diagonalized as follows, 
\begin{eqnarray}
&&T_{8}^{(m,0)}=\frac{1}{2\sqrt{3}}
diag
\left(m {\bf 1_{m+1}},(m-3){\bf 1},
(m-6){\bf 1},\cdots,-2m
\right), \cr 
&&T_{8}^{(m,n)}=\frac{1}{2\sqrt{3}}
diag
\left((m+2n){\bf 1_{m+1}},\cdots,
(-2m-n){\bf 1_{n+1}}
\right). 
\label{eigenvalueofT8}
\end{eqnarray}
The eigenvalues of $T_{8}^{(m,n)}$ are arranged 
in order of magnitude. 
We introduce {\it the south pole} 
by the point 
$T_{8}^{(m,0)}=-m/\sqrt{3}$ and 
$T_{8}^{(m,n)}=-(m+n/2)/\sqrt{3}$ $(m>n)$
in each representation. 
We can evaluate 
the quadratic Casimir of the $SU(2)$ 
algebra at the south pole, 
which is a subalgebra of the $SU(3)$ algebra 
\footnote{We used the formulae in 
Appendix \ref{sec:someformulaeofSU(3)} 
in this calculation.} 
as 
\begin{eqnarray}
&&\sum_{\mu=1}^{3}T_{\mu}^{(m,0)}T_{\mu}^{(m,0)}
=0, \cr
&&\sum_{\mu=1}^{3}T_{\mu}^{(m,n)}T_{\mu}^{(m,n)}
= \frac{n(n+2)}{4}. 
\label{S2atsouth}
\end{eqnarray}
This shows that the $(m,n)$ representation 
has a fuzzy two-sphere, which is given by 
the spin $n/2$ representation, 
at this point. 
The manifold which is constructed from the $(m,n)$ 
representation of $SU(3)$ is locally $CP^{2} \times S^{2}$, 
although it is not globally the case. 
We can summarize as follows. 
The $SU(3)$ algebra gives noncommutative manifolds 
whose isometry is $SU(3)$. 
There are two choices of stability group, 
$U(2)$ or $U(1) \times U(1)$. 
It must be noted that each manifold is a 
symplectic manifold 
(see \cite{aoyamamasuda} for example). 

The noncommutative coordinates 
can be realized by the guiding center coordinates 
on an ordinary commutative space in a magnetic monopole 
at the origin. 
The quantum Hall system on $CP^{2}$ was constructed in 
\cite{karabalinair}. 
Such a system is constructed in two ways; 
one is given by $U(1)$ gauge field background, 
and the other by combined $U(1)$ and $SU(2)$. 
These systems respectively correspond to 
the fuzzy spaces which are constructed by 
the $(m,0)$ or $(m,n)$ 
representation of $SU(3)$. 
Although the latter fuzzy space has an extra two-dimensional 
space, we may call it fuzzy 
$CP^{2}$ since it is realized by considering 
a commutative $CP^{2}$ and a magnetic monopole field. 

As has been shown in the previous section, 
the approach of a D$0$-brane to the surface of a 
fuzzy two-sphere induces instability. 
Analogous instability happens when a D$0$ brane 
approaches the surface of a fuzzy $CP^{2}$. 
The mass spectrum of the $0$-$4$ string is 
calculated in Appendix \ref{sec:instability}, 
and it is shown that tachyonic modes appear 
from the $0$-$4$ string. 
The fuzzy $CP^{2}$ basically has analogous 
instability to that of the fuzzy $S^{2}$. 

Before going to the next section, 
we will comment on the classical energy of these 
fuzzy spaces. 
It is found that the energy of the $(m,0)$ representation of 
$SU(3)$ is lower than that of the $(m^{\prime},n^{\prime})$ representation 
for fixed $N$ \cite{kitazawa}. 
We next compare the classical energy of the fuzzy two-sphere 
with that of the fuzzy $CP^{2}$. 
Since the energy of the irreducible representation 
is lower than that of the reducible representation, 
we consider only the irreducible representation. 
The energy is given by 
\begin{equation}
E_{S^{2}}=-\frac{\alpha^{4}}{6}j(j+1)(2j+1), 
\end{equation}
\begin{equation}
E_{CP^{2}}=-\frac{\alpha^{4}}{4}C_{2}(m,n)N(m,n),  
\label{energyCP^{2}}
\end{equation}
where $j=(N-1)/2$. 
We can find that $E_{S^{2}}$ is lower than 
$E_{CP^{2}}$ for fixed $N$. 
Therefore, if we consider some D$0$-branes 
under the $SU(3)$ invariant R-R four form background, 
a classical vacuum is realized by a fuzzy two-sphere 
configuration.


\section{Fuzzy sphere and fuzzy $CP^{2}$ from D$0$-branes}
\label{sec:D0branesinsu(3)}
\hspace{0.4cm}
In this section, we consider some D$0$-branes 
under the $SU(3)$ invariant R-R four form background. 
It can be expected that the fuzzy sphere and the fuzzy $CP^{2}$ 
are formed by the condensation of tachyons. 
(The last argument presented in the previous section 
showed that a configuration minimizing the potential 
is realized by the irreducible representation of $SU(2)$. )
The formation of these branes depends on the positions 
of the D$0$-branes. We will approach this analysis by considering 
several situations. 

\vspace{0.4cm}

We will first consider two D$0$-branes. 
The positions of them are 
\begin{equation}
\hat{x}_{i}=0 \hspace{0.2cm}(i\neq 3)
,\hspace{0.4cm}
\hat{x}_{3}=
\left( \begin{array}{c c}
  x_{3}^{(1)} & 0  \\  
  0 & x_{3}^{(2)}   \\
   \end{array} \right)\hspace{0.2cm}
( x_{3}^{(1)} -x_{3}^{(2)}\equiv \Delta >0), 
\end{equation}
and fields which appear from strings connecting these 
branes are presented by  
\begin{equation}
A_{\mu}=
\left( \begin{array}{c c c}
  0&a_{\mu}  \\  
  \bar{a}_{\mu} & 0   \\
   \end{array} \right). 
\end{equation}
For later convenience, we can calculate the 
mass terms in more general cases of  
$\Delta_{3}\neq 0$ and 
$\Delta_{8}\neq 0$:
\begin{eqnarray}
V_{2}&=&\frac{1}{2}Tr\left(
A_{\nu}[\hat{x}_{\mu},[\hat{x}_{\mu},
A_{\nu}]]
\right)
+2Tr\left(A_{\mu}
([\hat{x}_{\mu},\hat{x}_{\nu}]
-i\alpha f_{\mu\nu\lambda}\hat{x}_{\lambda}
)A_{\nu}\right), \cr
&=&
a_{\mu}(\delta_{\mu\nu} \Delta^{2}
+2i\alpha f_{\mu\nu \lambda}\Delta^{\lambda}
)\bar{a}_{\nu} \cr
&=&
(\Delta^{2}-2\alpha\Delta_{3})a_{+}^{(1)}\bar{a}_{-}^{(1)}
+(\Delta^{2}+2\alpha\Delta_{3})a_{-}^{(1)}\bar{a}_{+}^{(1)} \cr
&&+\Delta^{2}a_{3}\bar{a}_{3} \cr
&&+(\Delta^{2}-\alpha\Delta_{3}-\sqrt{3}\Delta_{8})
a_{+}^{(2)}\bar{a}_{-}^{(2)}
+(\Delta^{2}+\alpha\Delta_{3}+\sqrt{3}\Delta_{8})
a_{-}^{(2)}\bar{a}_{+}^{(2)} \cr
&&+(\Delta^{2}+\alpha\Delta_{3}-\sqrt{3}\Delta_{8})
a_{+}^{(3)}\bar{a}_{-}^{(3)}
+(\Delta^{2}-\alpha\Delta_{3}+\sqrt{3}\Delta_{8})
a_{-}^{(3)}\bar{a}_{+}^{(3)} \cr
&&+\Delta^{2}a_{8}\bar{a}_{8}, 
\label{masstermSU(3)}
\end{eqnarray}
where 
$\Delta^{2}\equiv (\Delta_{3})^{2}+(\Delta_{8})^{2}$ 
and 
\begin{equation}
a_{\pm}^{(1)}\equiv \frac{1}{\sqrt{2}}(a_{1}\pm ia_{2}), 
\hspace{0.2cm}
a_{\pm}^{(2)} \equiv \frac{1}{\sqrt{2}}(a_{4}\pm ia_{5}),
\hspace{0.2cm}
a_{\pm}^{(3)}\equiv \frac{1}{\sqrt{2}}(a_{6}\pm ia_{7}). 
\end{equation}
The $N=2$ case is related to $\Delta_{8}=0$ in the above mass 
terms. We then have  
\begin{eqnarray}
&&V_{2}=
(\Delta_{3}^{2}-2\alpha\Delta_{3})a_{+}^{(1)}\bar{a}_{-}^{(1)}
+(\Delta_{3}^{2}-\alpha\Delta_{3})
a_{+}^{(2)}\bar{a}_{-}^{(2)}
+(\Delta_{3}^{2}-\alpha\Delta_{3})
a_{-}^{(3)}\bar{a}_{+}^{(3)}. 
\end{eqnarray}
The other components are massless or massive. 
Let us first investigate the situation $0 <\Delta_{3} <\alpha$, 
where the three fields are tachyonic. 
After calculating the interaction 
terms, we can get the full potential,  
\begin{eqnarray}
&&V=(\Delta_{3}^{2}-2\alpha\Delta_{3})t_{1}^{2}
+(\Delta_{3}^{2}-\alpha\Delta_{3})t_{2}^{2}
+(\Delta_{3}^{2}-\alpha\Delta_{3})t_{3}^{2} \cr
&&\hspace{1cm}+t_{1}^{4}+t_{2}^{4}+t_{3}^{4}
+2t_{1}^{2}t_{2}^{2}+2t_{2}^{2}t_{3}^{2}+2t_{1}^{2}t_{3}^{2}, 
\end{eqnarray}
where $a_{+}^{(1)}=t_{1}e^{i\theta_{1}}$, 
$a_{+}^{(2)}=t_{2}e^{i\theta_{3}}$ and 
$a_{+}^{(3)}=t_{3}e^{i\theta_{3}}$. 
The point $t_{1}=t_{2}=t_{3}=0$ corresponds to 
an unstable extremum. On the other hand, a stable extremum 
is found to be 
\begin{equation}
t_{1}^{2}=-\frac{1}{2}m_{1}^{2}, 
\hspace{0.4cm}
t_{2}=t_{3}=0. 
\end{equation}
Although $t_{2}$ and 
$t_{3}$ are both zero, this point is stable. 
This situation is similar to the one encountered in 
equation (\ref{seemunstable}).  
The terms $\langle t_{1}^{2}\rangle t_{2}^{2}$ and 
$\langle t_{1}^{2} \rangle t_{3}^{2}$ can respectively 
give the mass of $t_{2}$ and 
$t_{3}$. 
The stability of this point is found from 
$d^{2}V/dt_{1}^{2}=-2m_{1}^{2}>0$, 
$d^{2}V/dt_{2}^{2}=d^{2}V/dt_{3}^{2}=2\alpha\Delta_{3}>0$. 
The configuration which is realized in this case 
is similar to 
a fuzzy ellipsoidal sphere (\ref{ellipsoidal}).
When $\Delta_{3}$ takes the interval 
$\alpha \le \Delta_{3} <2\alpha$, only $a_{+}^{(1)}$ can be 
tachyonic. The tachyon potential is given by 
\begin{equation}
V=(\Delta_{3}^{2}-2\Delta_{3})(a_{+}^{(1)}\bar{a}_{-}^{(1)})
+(a_{+}^{(1)}\bar{a}_{-}^{(1)})^{2}. 
\end{equation}
When $\Delta_{3}$ is $\alpha$, 
a fuzzy sphere, which is constructed 
from the spin-$1/2$ representation of $SU(2)$, 
is obtained after tachyon condensation. 
This is the minimum of this potential.

\vspace{0.4cm}

The next analysis is devoted to three D$0$-branes. 
In this case, we can expect the formation both of the fuzzy 
$S^{2}$ and the fuzzy $CP^{2}$. 
If we consider the following background, 
\begin{equation}
\hat{x}_{3}=\alpha L_{3}^{(j=1)},\hspace{0.3cm} 
\hat{x}_{i}=0\hspace{0.1cm}(i\neq3), 
\end{equation}
we would obtain the spin-$1$ representation of $SU(2)$. 
Since this calculation is almost analogous to other cases, 
we omit the detail of it. 
The value of the classical potential for this solution 
is $V=-\alpha^{4}$. 

We are next concerned with the following background: 
\begin{equation}
\hat{x}_{3}=\frac{\alpha}{2}
\lambda_{3}, \hspace{0.3cm} 
\hat{x}_{8}=\frac{\alpha}{2}
\lambda_{8},\hspace{0.3cm}
\hat{x}_{i}=0\hspace{0.1cm}(i\neq 3,8).  
\end{equation}
The distances between these D$0$-branes are 
\begin{eqnarray}
&&\Delta_{(12)}^{3}=\alpha, 
\hspace{0.4cm}
\Delta_{(23)}^{3}=-\frac{1}{2}\alpha,
\hspace{0.4cm} 
\Delta_{(13)}^{3}=\frac{1}{2}\alpha, \cr
&&\Delta_{(12)}^{8}= 0, 
\hspace{0.4cm}
\Delta_{(23)}^{8}=\frac{\sqrt{3}}{2}\alpha,
\hspace{0.4cm}
\Delta_{(13)}^{8}=\frac{\sqrt{3}}{2}\alpha. 
\label{distanceSU(3)N=3}
\end{eqnarray}
We parameterize the fluctuations around this solution as 
\begin{equation}
A_{\mu}=
\left( \begin{array}{c c c}
  0&a_{\mu} &b_{\mu} \\  
  \bar{a}_{\mu} & 0  & c_{\mu} \\
  \bar{b}_{\mu} &\ \bar{c}_{\mu} & 0 \\
   \end{array} \right). 
\end{equation}
The mass terms can be easily calculated 
by substituting (\ref{distanceSU(3)N=3}) 
into (\ref{masstermSU(3)}) to give 
\begin{equation}
V_{2}/\alpha^{2}=-a_{+}^{(1)}\bar{a}_{-}^{(1)}
-b_{+}^{(2)}\bar{b}_{-}^{(2)}-c_{+}^{(3)}\bar{c}_{-}^{(3)}, 
\end{equation}
where we have dropped non-tachyonic fields. 
The interaction terms for the tachyonic fields 
are calculated as follows: 
\begin{eqnarray}
V_{3}&=&-\sqrt{2}\alpha 
(a_{+}^{(1)}\bar{b}_{+}^{(2)}c_{+}^{(3)}
+\bar{a}_{+}^{(1)}b_{+}^{(2)}\bar{c}_{+}^{(3)}), \cr
V_{4}&=&
(a_{+}^{(1)}\bar{a}_{-}^{(1)})^{2}
+(b_{+}^{(2)}\bar{b}_{-}^{(2)})^{2}
+(c_{+}^{(3)}\bar{c}_{-}^{(3)})^{2} \cr
&&
+(a_{+}^{(1)}\bar{a}_{-}^{(1)})(b_{+}^{(2)}\bar{b}_{-}^{(2)})
+(b_{+}^{(2)}\bar{b}_{-}^{(2)})(c_{+}^{(3)}\bar{c}_{-}^{(3)})
+(c_{+}^{(3)}\bar{c}_{-}^{(3)})(a_{+}^{(1)}\bar{a}_{-}^{(1)}). 
\end{eqnarray}
Note that $V_{3}$ is not zero, while it is zero 
in the $SU(2)$ background. 
Defining 
\begin{equation}
a_{+}^{(1)}\equiv t_{1}e^{i\theta_{1}},\hspace{0.4cm}
b_{+}^{(2)}\equiv t_{2}e^{i\theta_{2}},\hspace{0.4cm}
c_{+}^{(3)}\equiv t_{3}e^{i\theta_{3}}, 
\end{equation}
the extrema satisfy 
\begin{eqnarray}
&&-2t_{1}+4t_{1}^{3}+2t_{1}t_{2}^{2}+2t_{1}t_{3}^{2}
-2\sqrt{2}t_{2}t_{3}\cos (\theta_{1}-\theta_{2}+\theta_{3})
=0, \cr
&&-2t_{2}+4t_{2}^{3}+2t_{2}t_{1}^{2}+2t_{2}t_{3}^{2}
-2\sqrt{2}t_{1}t_{3}\cos (\theta_{1}-\theta_{2}+\theta_{3})
=0, \cr
&&-2t_{3}+4t_{3}^{3}+2t_{3}t_{1}^{2}+2t_{3}t_{2}^{2}
-2\sqrt{2}t_{1}t_{2}\cos (\theta_{1}-\theta_{2}+\theta_{3})
=0. 
\end{eqnarray}
$t_{1}=t_{2}=t_{3}=0$ is a trivial solution, representing 
an unstable extremum. 
The stable solution we are searching for is found to be 
$t_{1}=t_{2}=t_{3}=1/\sqrt{2}$ and  
$\theta_{1}=\theta_{2}=\theta_{3}=0$. 
This solution provides the $(1,0)$ representation of 
$SU(3)$, that is to say, 
$X_{\mu}$ are represented as $\alpha \lambda_{\mu}/2$. 
The value of the classical potential for this solution 
is $V=-\alpha^{4}$. 
For the $N=3$ case, we could obtain two noncommutative solutions. 
Although we can get other noncommutative solutions 
by changing the positions of the D$0$-branes, such noncommutative 
solutions do not minimize the classical potential. 
The classical energy of these two branes which we have obtained 
coincides accidentally. 
In general, the fuzzy two-sphere solution has lower classical energy 
than the fuzzy $CP^{2}$ solution. 
If we consider consider quantum effects, 
the transition between these solutions is expected to be seen. 

\vspace{0.4cm}

It was shown in section \ref{sec:fuzzyspherefromD0} that 
a complex tachyon appears 
when a D$0$-brane approaches the surface 
of a spherical D$2$-brane, and that  
condensation of the tachyon produces 
a spherical D$2$-brane with a larger radius. 
If we embed a system composed of a D$0$-brane 
and a D$2$-brane into eight-dimensional space 
under the $SU(3)$ invariant R-R four form background, 
we can observe a more interesting phenomenon. 
The condensation of tachyons appearing 
in this system produces 
a D$4$-brane which is constructed 
from the $SU(3)$ algebra.  
To observe such a phenomenon, 
let us start with the following classical solution: 
\begin{eqnarray}
&&\hat{x}_{i}=\frac{\alpha}{2}\lambda_{i}=
\frac{\alpha}{2}
\left( \begin{array}{c c}
   \sigma_{i}  & 0  \\
   0& 0 \\
   \end{array} \right) 
   \hspace{0.1cm}(i=1,2,3), 
   \hspace{0.4cm} \cr
&&\hat{x}_{8}=d\frac{\alpha}{2}\lambda_{8}
=\frac{\alpha}{2\sqrt{3}}
\left( \begin{array}{c c}
   d{\bf 1}_{2}  & 0  \\
   0& -2d \\
   \end{array} \right). 
\end{eqnarray}
The other coordinates are zero. 
The upper left part of the matrices 
represents a fuzzy two-sphere whose position is 
$(x_{3},x_{8})=(0,\alpha d/2\sqrt{3})$, and 
the lower right part 
a D$0$-brane whose position is $(0, -\alpha d/\sqrt{3})$. 
If $d$ is infinity, 
these two branes are far away in the eighth direction, and 
there do not appear any tachyonic modes from strings 
connecting these two branes. When $d$ takes a value, 
we can show that tachyonic fields appear. 
The mass term is evaluated as 
\begin{eqnarray}
&&V_{2}/\alpha^{2}=
\frac{3}{4}(d^{2}+1) v_{\mu}^{\dagger}v_{\mu}\cr
&&+\frac{3}{2}id(v_{5}^{\dagger}v_{4}-v_{4}^{\dagger}v_{5}
+v_{7}^{\dagger}v_{6}-v_{6}^{\dagger}v_{7})
+\frac{1}{2}id(v_{5}^{\dagger}\sigma_{3}v_{4}-v_{4}^{\dagger}\sigma_{3}v_{5}
+v_{7}^{\dagger}\sigma_{3}v_{6}-v_{6}^{\dagger}\sigma_{3}v_{7}) \cr
&&-\frac{1}{2}i(v_{4}^{\dagger}\sigma_{1}v_{7}-v_{7}^{\dagger}\sigma_{1}v_{4}
-v_{5}^{\dagger}\sigma_{1}v_{6}+v_{6}^{\dagger}\sigma_{1}v_{5}
+v_{4}^{\dagger}\sigma_{2}v_{6}-v_{6}^{\dagger}\sigma_{2}v_{4}
+v_{5}^{\dagger}\sigma_{2}v_{7}-v_{7}^{\dagger}\sigma_{2}v_{5}),  
\end{eqnarray}
where $v_{\mu}$ and $v_{\mu}^{\dagger}$ are off-diagonal 
fluctuations. 
It can be shown that the tachyonic modes do not appear from 
$v_{i}\hspace{0.1cm}(i=1,2,3,8)$, the 
second component of $v_{4}$, $v_{5}$ 
and the first component of $v_{6}$, $v_{7}$. Therefore, 
we make them zero. In the remainder of this calculation, 
$v_{4}$ and $v_{5}$ denote their first components, and 
$v_{6}$ and $v_{7}$ their second components. 
The mass terms for $v_{4}$, $v_{5}$, $v_{6}$ and $v_{7}$ 
are diagonalized as follows, 
\begin{eqnarray}
&&V_{2}/\alpha^{2}=
\left(\frac{3}{4}d^{2}-2d-\frac{1}{4}\right)
\frac{1}{2}(v_{+}^{(2)}+v_{+}^{(3)})
(\bar{v}_{-}^{(2)}+\bar{v}_{-}^{(3)}) \cr
&&\hspace{1cm}
+\left(\frac{3}{4}d^{2}-2d+\frac{7}{4}\right)
\frac{1}{2}(v_{+}^{(2)}-v_{+}^{(3)})
(\bar{v}_{-}^{(2)}-\bar{v}_{-}^{(3)})\cr
&&\hspace{1cm}
+\left(\frac{3}{4}d^{2}+2d+\frac{3}{4}\right) 
\frac{1}{2}v_{-}^{(2)}
\bar{v}_{+}^{(2)} 
+\left(\frac{3}{4}d^{2}+2d+\frac{3}{4}\right)
\frac{1}{2}v_{-}^{(3)}
\bar{v}_{+}^{(3)}, 
\label{massD0D2D4}
\end{eqnarray}
where 
\begin{equation}
v_{\pm}^{(2)}=\frac{1}{\sqrt{2}}(v_{4}\pm v_{5}), 
\hspace{0.4cm}
v_{\pm}^{(3)}=\frac{1}{\sqrt{2}}(v_{6}\pm v_{7}). 
\end{equation}
Only the first term can be a tachyonic mass term when 
$d$ takes the following interval,
\begin{eqnarray}
\frac{4}{3}-\frac{\sqrt{19}}{3}<d<
\frac{4}{3}+\frac{\sqrt{19}}{3}. 
\end{eqnarray}
This instability is different from 
one we confronted in the section \ref{sec:fuzzyspherefromD0} 
or Appendix \ref{sec:instability} which happens 
when a D$0$-brane approaches the 
surface of a fuzzy space. 
This instability happens when the distance between two branes 
in the eighth direction takes a value. 
It is interesting to consider $d=1$. 
Only the first term in (\ref{massD0D2D4}) 
is relevant to this case. 
The full tachyon potential in this case is 
\begin{equation}
V=-3t^{2}+3t^{4}
\end{equation}
where $v_{+}^{(2)}=v_{+}^{(3)}=t e^{i\theta}$. 
The end point of tachyon condensation is given by 
$t=1/\sqrt{2}$, where the fuzzy $CP^{2}$ 
given by the $(1,0)$ representation of 
$SU(3)$ is realized. 


\section{Summary and discussions}
\label{sec:summaryanddiscussions}
\hspace{0.4cm}
In this work, we studied tachyon condensation 
in the process of the Myers effect by using the low energy 
effective Yang-Mills action. 
D$0$-branes under the R-R field strength background 
are unstable, and we can expect that this system 
can become stable by the condensation of tachyon. 
We have considered two kinds of R-R field strength, 
one being given by the structure constant of $SU(2)$, and 
the other, by that of $SU(3)$. 
The characteristic configurations of these cases are 
the fuzzy sphere and the fuzzy $CP^{2}$ respectively. 

In general, a classical solution given by 
a reducible representation is unstable, and 
it is the irreducible representation that minimizes 
the classical potential. 
We explicitly confirmed that the instability of 
the reducible representations is manifested by the existence of 
tachyonic fields, 
and that the irreducible representation is realized 
after the condensation of the tachyons. 
A big difference between $SU(2)$ and $SU(3)$ is 
that the size of their matrices is restricted in the latter case 
as in (\ref{sizesu(3)matrix}), 
while it is not in the former case. 
For the $SU(2)$ invariant background case, 
the only stable configuration is a fuzzy sphere 
whose coordinate is given by the irreducible representation of 
$SU(2)$. The fact that the reducible representation of $SU(2)$ 
rolls down to the irreducible representation of $SU(2)$ 
does not depend on the dimension of the representation. 
On the other hand, this is not the case 
when we consider the reducible representation of $SU(3)$. 
Although the formation of the irreducible representation of $SU(3)$
from the reducible representation of $SU(3)$ is not always seen, 
tachyonic instability always appears when 
the reducible representation is considered. 
\footnote{
It is shown in the Appendix \ref{sec:instability} that 
a reducible representation which is composed of 
a D$0$-brane and a fuzzy $CP^{2}$ 
is unstable due to the existence of a complex 
tachyonic field. 
This proof can be extended to general reducible representations 
of $SU(3)$. 
We should not overlook that 
this discussion is independent of the size of the representation. 
}
Since runaway behavior never happens from the shape of 
the potential, a noncommutative 
solution can expected to appear. 
\footnote{
Such an idea has already been applied in \cite{boer}. }

An interesting property of matrix models or noncommutative 
gauge theories is background independence \cite{seiberg}. 
The matrix model variables are $N\times N$ matrices $X_{\mu}$. 
Different D-branes appear as different classical solutions 
of matrix models. The fields on the branes can be described by 
expanding $X_{\mu}$ around the classical solutions 
as given in equation (\ref{expansion}). 
We have treated in this study a simple matrix model action 
with a Chern-Simons term. 
Different curved D-branes arose as classical solutions, 
and the condensation of tachyons led to the transition between 
different D-branes. 

As shown in the section \ref{sec:D0branesinsu(3)}, 
a fuzzy $CP^{2}$ is formed from a D$0$-brane and 
a fuzzy $S^{2}$ through the condensation of tachyon. 
A similar phenomenon such as the formation of the fuzzy 
$CP^{2}$ from the fuzzy $CP^{2}$ and the fuzzy $S^{2}$ 
can be seen. The important thing leading to these phenomena 
is that the matrix representation of $SU(2)$ is included 
in that of $SU(3)$. In general, we can say that 
D-branes whose coordinates are given by 
higher dimensional Lie-algebra 
shows interesting dynamics. 
In this sense, it is also an interesting problem 
to study the higher dimensional fuzzy sphere 
\cite{horamgoolam,kimura}
in the context of the tachyon condensation.

\vspace{1.0cm}
\begin{center}
{\bf Acknowledgments}
\end{center}
\hspace{0.4cm}
I acknowledge helpful discussions with T. Masuda. 
I also thank colleagues at RIKEN for 
some discussions. 

\renewcommand{\theequation}{\Alph{section}.\arabic{equation}}

\appendix

\section{$(m,n)$ representation of $SU(3)$}
\setcounter{equation}{0} 
\label{sec:someformulaeofSU(3)}
\hspace{0.4cm}
In this Appendix, we construct some formulae 
which are relevant to the $(m,n)$ representation 
of $SU(3)$. 
$T_{\mu}^{(m,n)}$ denotes the generator 
of $(m,n)$ representation, being constructed as 
\begin{eqnarray}
T_{\mu}^{(m,n)}
&=&\left(t_{\mu} \otimes 1 \otimes \cdots \otimes 1 
+1 \otimes t_{\mu} \otimes  \cdots \otimes 1 
+\cdots + 
1 \otimes \cdots \otimes t_{\mu} \otimes \cdots \otimes 1 
\right. \cr 
&& \left. 
+ 1\otimes \cdots \otimes 1 \otimes s_{\mu} \otimes 
\cdots \otimes1 
+ \cdots +
1\otimes \cdots \otimes 1 \otimes s_{\mu}
\right), 
\end{eqnarray}
where $t_{\mu}$ and $s_{\mu}=-t_{\mu}^{*}$ are the generators of 
fundamental and anti-fundamental representation 
respectively. $t_{\mu}$ is written as $\lambda_{\mu}/2$ 
using the Gell-Mann matrix. 
$T_{\mu}^{(m,n)}$ is assumed to act on the following state, 
\begin{eqnarray}
\mid a_{1},a_{2},\ldots,a_{m},\bar{a}_{m+1},\ldots ,
\bar{a}_{m+n} \hspace{0.1cm}\rangle. 
\end{eqnarray}
where this state has the same symmetry as the 
Young diagram corresponding to the $(m,n)$ representation of 
$SU(3)$. 
The dimension of this sate is given by (\ref{sizesu(3)matrix}). 

We provide some useful formulae, 
\begin{eqnarray}
&&\sum_{\mu=1}^{8}\left(t_{\mu} \otimes t_{\mu} \right)
=\frac{1}{3}\left(1 \otimes 1 \right),  \cr 
&&\sum_{\mu=1}^{8}\left(t_{\mu}t_{\nu} \otimes t_{\mu} \right)
=\frac{1}{3}\left(t_{\nu} \otimes 1 \right), 
\hspace{0.4cm}
\sum_{\mu=1}^{8}\left(t_{\nu}t_{\mu} \otimes t_{\mu} \right)
=\frac{1}{3}\left(t_{\nu} \otimes 1 \right). 
\end{eqnarray}
These are easily proved by using 
\begin{equation}
\sum_{\mu=1}^{8}(t_{\mu})_{ij}(t_{\mu})_{kl}=
\frac{1}{2}\delta_{il}\delta_{jk}-
\frac{1}{6}\delta_{ij}\delta_{kl}. 
\end{equation}
We also provide 
\begin{equation}
\sum_{\mu=1}^{8}\left(t_{\mu} \otimes s_{\mu} \right)
=\frac{1}{6}\left(1 \otimes 1 \right).  
\end{equation}
The quadratic Casimir is easily calculated by using 
the above formulae, 
\begin{equation}
C_{2}(m,n)=T_{\mu}^{(m,n)}T_{\mu}^{(m,n)}=
\frac{1}{3}\left(
m^{2}+3m+n^{2}+3n+mn
\right)1_{N}. 
\label{casimirsu3}
\end{equation}


\section{Instability of curved branes}
\label{sec:instability}
\setcounter{equation}{0} 
\hspace{0.4cm}
In this Appendix, we observe the 
instability of reducible representations. 
We consider a system 
of a fuzzy brane and a D$0$-brane 
as a simple example of reducible representations. 
Tachyonic modes appearing from strings 
connecting a fuzzy brane and a D$0$-brane 
manifest the instability of this system. 

We first show the 
instability of D$0$-D$2$ system. 
The detail of 
this calculation is found in \cite{hashimotokrasnov}. 
The configuration of a D$0$-brane 
and a spherical D$2$-brane is represented by 
\begin{equation}
L_{\mu}= 
\left( \begin{array}{c c }
  c_{\mu}/\alpha & 0\\  
  0 & L_{\mu}^{(j)}   \\
   \end{array} \right), 
\end{equation}
where $c_{\mu}$ is a number and 
$L_{\mu}^{(j)}$ is the spin $j \equiv (N-2)/2$ 
representation of $SU(2)$. 
A complex tachyon appears from off diagonal parts 
of the fluctuations, 
\begin{equation}
A_{\mu}= 
\left( \begin{array}{c c }
  0 & v_{\mu}\\  
  v_{\mu}^{\dagger} & 0   \\
   \end{array} \right),  
\end{equation}
where $v_{\mu}$ are $1\times (N-1)$ matrices. 
Mass terms for $v_{\mu}$ are calculated as 
\begin{eqnarray}
V_{2}/\alpha^{2}&=&\left(c^{2}-2c-2cm+ j(j+1)\right)
v_{+}v_{-}^{\dagger}+\left(c^{2}+2c-2cm+ j(j+1)\right)
v_{-}v_{+}^{\dagger} \cr
&&+\left(c^{2}-2cm+ j(j+1)\right)
v_{3}v_{3}^{\dagger},  
\end{eqnarray}
where $m$ runs over $-j$ to $j$. 
We have used the rotation symmetry to 
fix $c_{\mu}$ as $c\delta_{\mu 3}$, 
which does not lose generality. 
If we suppose $c \ge 0$, only 
an $m=j$ component of 
$v_{+}$ (or $v_{-}^{\dagger}$) 
becomes tachyonic  
when 
the position of the D$0$-brane 
takes the following interval,
\begin{equation}
j+1 
-\sqrt{j+1}
<\frac{c}{\alpha}<
j+1 
+\sqrt{j+1}.
\label{unstableregion}
\end{equation}
This shows that 
a complex tachyon appears when  
a D$0$-brane is close to the surface 
of a spherical D$2$-brane.
This manifests the instability of the system, 
and after the tachyon condensation this system 
is expected to become 
the irreducible representation which 
is given by the spin $(N-1)/2$ representation 
of $SU(2)$. 

\vspace{0.4cm}

We next show the instability of 
D$0$-D$4$ system where a D$4$-brane forms fuzzy $CP^{2}$. 
This calculation is analogous to the previous D$0$-D$2$ system. 
A classical solution representing such a system is given by 
\begin{equation}
J_{\mu}= 
\left( \begin{array}{c c }
  c_{\mu}/\alpha & 0\\  
  0 & T_{\mu}^{(m,n)}   \\
   \end{array} \right), 
\end{equation}
where $c_{\mu}$ are the positions of a D$0$-brane 
and $T_{\mu}^{(m,n)}$ $(m\ge n)$ are the $(m,n)$ representation of $SU(3)$. 
The size of the matrices $J_{\mu}$ is $N+1$ 
where $N$ is given by (\ref{sizesu(3)matrix}). 
The mass term is calculated as 
\begin{equation}
V_{2}/\alpha^{2}=v_{\nu}^{\dagger}
\left(T_{\mu}^{(m,n)}-c_{\mu}1_{N}\right)
\left(T_{\mu}^{(m,n)}-c_{\mu}1_{N}\right)v_{\nu}
+2if_{\mu\nu\lambda}
v_{\mu}^{\dagger}v_{\nu}c_{\lambda}. 
\end{equation}

Let us consider the case 
$c_{\mu}=(0,0,c_{1},0,0,0,0,c_{2})$. 
The above mass terms become 
\begin{eqnarray}
V_{2}/\alpha^{2}&=&
\left(c_{1}^{2}+c_{2}^{2}
-2(c_{1} t_{3}+c_{2}t_{8})+C_{2}(m,n)\right)
v_{\mu}^{\dagger}v_{\mu}\cr 
&&+2ic_{1}
\left(
v_{1}^{\dagger}v_{2}-v_{2}^{\dagger}v_{1}\right)
+ic_{1}
\left(
v_{4}^{\dagger}v_{5}-v_{5}^{\dagger}v_{4}
-v_{6}^{\dagger}v_{7}+v_{7}^{\dagger}v_{6}
\right) \cr
&&+\sqrt{3}ic_{2}
\left(
v_{4}^{\dagger}v_{5}-v_{5}^{\dagger}v_{4}
+v_{6}^{\dagger}v_{7}-v_{7}^{\dagger}v_{6}
\right). 
\end{eqnarray}
In this expression, $t_{3}$ and $t_{8}$ denote 
eigenvalues of $T_{3}^{(m,n)}$ and $T_{8}^{(m,n)}$ respectively. 
These mass terms are diagonalized as 
\begin{eqnarray}
V_{2}/\alpha^{2}&=&
m_{1}^{2}
\left(v_{3}^{\dagger}v_{3}+v_{8}^{\dagger}v_{8}\right)
+m_{2+}^{2}v_{-}^{(1)\dagger}v_{+}^{(1)}
+m_{2-}^{2}v_{+}^{(1)\dagger}v_{-}^{(1)} \cr
&&+m_{3+}^{2}v_{-}^{(2)\dagger}v_{+}^{(2)}
+m_{3-}^{2}v_{+}^{(2)\dagger}v_{-}^{(2)}
+m_{4+}^{2}v_{-}^{(3)\dagger}v_{+}^{(3)}
+m_{4-}^{2}v_{+}^{(3)\dagger}v_{-}^{(3)},
\end{eqnarray}
where the mass is given by 
\begin{eqnarray}
&& m_{1}^{2}=c_{1}^{2}+c_{2}^{2}
-2(c_{1} t_{3}+c_{2}t_{8})+C_{2}, 
\hspace{2.8cm}
\cr
&& m_{2\pm}^{2}=c_{1}^{2}+c_{2}^{2}
-2(c_{1} t_{3}+c_{2}t_{8})+C_{2}\pm 2c_{1},
\hspace{1.8cm}
\cr
&& m_{3\pm}^{2}=c_{1}^{2}+c_{2}^{2}
-2(c_{1} t_{3}+c_{2}t_{8})+C_{2}
\pm (c_{1}+\sqrt{3}c_{2}), 
\hspace{0.4cm}
\cr 
&&
m_{4\pm}^{2}=c_{1}^{2}+c_{2}^{2}
-2(c_{1} t_{3}+c_{2}t_{8})+C_{2}
\pm (-c_{1}+\sqrt{3}c_{2}).
\end{eqnarray}
We find after careful consideration that 
$m_{1}^{2}$, $m_{2\pm}^{2}$, $m_{3-}^{2}$ and $m_{4-}^{2}$ 
are always positive. 
We can show that 
$m_{3+}^{2}$ and $m_{4+}^{2}$
can be negative only if 
$(t_{8},t_{3})=(-(m+n/2)/\sqrt{3},-n/2)$ 
and $(-(m+n/2)/\sqrt{3},n/2)$ respectively. 
The value of $t_{8}$ corresponds to the minimal value of 
$T_{8}^{(m,n)}$ (see (\ref{eigenvalueofT8})). 
As calculated in (\ref{S2atsouth}), 
a fuzzy two-sphere is located in this point, which 
is given by the spin $n/2$ representation of $SU(2)$. 
Therefore 
$t_{3}=+n/2$,$-n/2$ are the maximal and minimal 
values of $T_{3}^{(m,n)}$. 
The inequality we need to solve is 
\begin{eqnarray}
m_{3+,4+}^{2}&=& c_{1}^{2}-(2t_{3}\mp1)c_{1}
+c_{2}^{2}-(2t_{8}-\sqrt{3})c_{2}+C_{2} \cr
&=&\left(c_{1}\pm\frac{n+1}{2}\right)^{2}
+\left(c_{2}+\left(\frac{m}{3}+\frac{n}{6}+\frac{1}{2}\right)
\sqrt{3}\right)
-1 <0, 
\end{eqnarray}
where we have substituted 
$t_{8}=-(m+n/2)/\sqrt{3}$ and $t_{3}=\mp n/2$. 
The expressions of 
$m_{3+}^{2}$ and $m_{4+}^{2}$ are exchanged by changing 
the sign of $c_{1}$. 
We first regard this inequality as 
the second order one of $c_{1}$, it being solved as 
\begin{equation}
\mp\frac{n+1}{2}-\sqrt{D}<\frac{c_{1}}{\alpha}<
\mp \frac{n+1}{2}+\sqrt{D}, 
\label{c1inequality} 
\end{equation}
where $D$ satisfies 
\begin{equation}
0 < D \le \frac{n}{2}+\frac{7}{4}. 
\end{equation}
The condition $D > 0$ gives 
\begin{equation}
-\frac{m+\frac{n}{2}}{\sqrt{3}}-\frac{\sqrt{3}}{2}
-\sqrt{\frac{n}{2}+\frac{7}{4}}<\frac{c_{2}}{\alpha}<
-\frac{m+\frac{n}{2}}{\sqrt{3}}-\frac{\sqrt{3}}{2}
+\sqrt{\frac{n}{2}+\frac{7}{4}}. 
\end{equation}
We have shown that 
a complex tachyon appears when a D$0$-brane approaches 
the surface of a fuzzy $CP^{2}$. 
We would like to emphasize that 
the equation (\ref{c1inequality}) reflects the fact that 
the fuzzy $CP^{2}$ has the extension along 
a fuzzy two-sphere (see (\ref{S2atsouth})).

\end{document}